\documentclass[conference]{IEEEtran}
\ifCLASSINFOpdf
\else
\fi
\usepackage[cmex10]{amsmath}
\usepackage{fixltx2e}

\usepackage{stfloats}

\usepackage{url}

\usepackage{amssymb}
\usepackage{amsmath}
\usepackage{mathdots}

\usepackage{color}

\def\mymedskip{\vskip\medskipamount}
\def\mymedbreak{\par \ifdim\lastskip<\medskipamount
  \removelastskip \penalty-100 \mymedskip \fi}
\def\myaftermedspace{\par \ifdim\lastskip<\medskipamount
  \removelastskip \penalty55\mymedskip\fi}
\newcommand{\eop}{{\unskip\nobreak\hfil\penalty50
          \hskip2em\hbox{}\nobreak\hfil$\Box$
          \parfillskip=0pt \finalhyphendemerits=0 \par}}
\newenvironment{proof}%
{\mymedbreak{\noindent\bf Proof:\enspace}}{\eop\myaftermedspace}
\newenvironment{proofone}%
{\mymedbreak{\noindent\bf Proof 1:\enspace}}{\eop\myaftermedspace}
\newenvironment{prooftwo}%
{\mymedbreak{\noindent\bf Proof 2:\enspace}}{\eop\myaftermedspace}
{\mymedbreak{\noindent\bf Proof of Theorem #1:\enspace}}{\eop\myaftermedspace}
{\mymedbreak{\noindent\bf Proof of Lemma #1:\enspace}}{\eop\myaftermedspace}
{\mymedbreak{\noindent\bf Proof of Proposition #1:\enspace}}{\eop\myaftermedspace}
{\mymedbreak\noindent{\bf Remark:}%
\enspace\rm}%
{\myaftermedspace}
\newtheorem{teor}{Theorem}[section]
\newtheorem{defi}[teor]{Definition}
\newtheorem{examp}[teor]{Example}
\newtheorem{lem}[teor]{Lemma}
\newtheorem{cor}[teor]{Corollary}
\newtheorem{con}[teor]{Conjecture}
\newtheorem{prop}[teor]{Proposition}
\newtheorem{rem}[teor]{Remark}
\newcommand{\beq}{\begin{equation}}
\newcommand{\eeq}{\end{equation}}
\newcommand{\beql}[1]{\begin{equation} \label{#1}}
\newcommand{\eeql}{\end{equation}}
\newcommand{\beqa}{\begin{eqnarray*}}
\newcommand{\eeqa}{\end{eqnarray*}}
\newcommand{\beqal}[1]{\begin{eqnarray} \label{#1}}
\newcommand{\eeqal}{\end{eqnarray}}
\newcommand{\beqan}{\begin{eqnarray}}
\newcommand{\eeqan}{\end{eqnarray}}
\newcommand{\bpf}{\begin{proof}}
\newcommand{\epf}{\end{proof}}
\newcommand{\bpfone}{\begin{proofone}}
\newcommand{\epfone}{\end{proofone}}
\newcommand{\bpftwo}{\begin{prooftwo}}
\newcommand{\epftwo}{\end{prooftwo}}

\newcommand{\cU}{{\cal U}}

\newcommand{\ga}{\alpha}
\newcommand{\gb}{\beta}
\newcommand{\gc}{\gamma}

\newcommand{\bab}{\begin{abstract}}
\newcommand{\eab}{\end{abstract}}
\newcommand{\bke}{\begin{keywords}}
\newcommand{\eke}{\end{keywords}}

\newcommand{\btm}[1]{\begin{teor} \label{#1}}
\newcommand{\etm}{\end{teor}}
\newcommand{\btmn}[2]{\begin{teor}[#1] \label{#2}}
\newcommand{\etmn}{\end{teor}}
\newcommand{\ble}[1]{\begin{lem} \label{#1}}
\newcommand{\ele}{\end{lem}}
\newcommand{\bLe}[1]{\begin{Lemma} \label{#1}}
\newcommand{\eLe}{\end{Lemma}}
\newcommand{\bpn}[1]{\begin{prop} \label{#1}}
\newcommand{\epn}{\end{prop}}
\newcommand{\bex}[1]{\begin{examp} \label{#1}}
\newcommand{\eex}{\end{examp}}
\newcommand{\bde}[1]{\begin{defi} \label{#1}}
\newcommand{\ede}{\end{defi}}
\newcommand{\bco}[1]{\begin{cor} \label{#1}}
\newcommand{\eco}{\end{cor}}
\newcommand{\bcorn}[2]{\begin{cor}[#1] \label{#1}}
\newcommand{\ecorn}{\end{cor}}
\newcommand{\bcon}[1]{\begin{con} \label{#1}}
\newcommand{\econ}{\end{con}}
\newcommand{\bre}[1]{\begin{rem} \label{#1}}
\newcommand{\ere}{\end{rem}}

\newcommand{\bbF}{\mathbb{F}}

\newcounter{question_number}
\newenvironment{question}{\addtocounter{question_number}{1}\noindent{\bf Question \arabic{question_number}}}{\myaftermedspace}

\newenvironment{solution}{\noindent {\bf Solution:} \enspace}{\eop\myaftermedspace}

\newenvironment{hint}{\noindent {\bf Hint:} \enspace}{\eop\myaftermedspace}

\newenvironment{multisolution}[1]{\noindent {\bf Solution #1:} \enspace}{\eop\myaftermedspace}

\newcommand{\bqu}{\begin{question}}
\newcommand{\equ}{\end{question}}
\newcommand{\bs}{\begin{solution}}
\newcommand{\es}{\end{solution}}
\newcommand{\bh}{\begin{hint}}
\newcommand{\eh}{\end{hint}}
\newcommand{\bms}[1]{\begin{multisolution}{#1}}
\newcommand{\ems}{\end{multisolution}}

\renewcommand{\bpf}{\begin{IEEEproof}}
\renewcommand{\epf}{\end{IIEEEproof}}
\renewcommand{\IEEEQED}{\IEEEQEDopen}

\begin{document}
%
\title{Storage codes -- coding rate and repair locality}

\author{\IEEEauthorblockN{Henk D. L.\ Hollmann}
\IEEEauthorblockA{%
Division of Mathematical
Sciences, School of Physical and Mathematical Sciences,\\Nanyang Technological
University, 
Singapore
\\Email: {\tt
Henk.Hollmann@ntu.edu.sg}%
}
}


%

\IEEEspecialpapernotice{(Invited Paper)}

\maketitle

\begin{abstract}
The {\em repair locality\/} of a distributed storage code is the maximum number of nodes that ever needs to be contacted during the repair of a failed node. Having small repair locality is desirable, since it is proportional to the number of disk accesses during repair. 
However, recent publications show that small repair locality comes with a penalty in terms of code distance or storage overhead  if exact repair is required.

Here, we first review some of the main results on storage codes under various repair regimes and discuss the recent work on possible 
(information-theoretical)
trade-offs between repair locality and other code parameters  like storage overhead and 
code distance, under the exact repair regime.

Then we present some new information theoretical lower bounds on the storage overhead as a function of the repair locality, valid for all common coding and repair models.
In particular, we show that if each of the $n$ nodes in a distributed storage system  has storage capacity $\ga$ and if, at any time, a failed node can be {\em functionally\/} repaired by contacting {\em some\/} 
set of~$r$
nodes 
(which may depend on the actual state of the system) 
and downloading an amount $\gb$  of data from each, then in the extreme cases where $\ga=\gb$ or $\ga = r\gb$, the maximal coding rate is at most $r/(r+1)$ or $1/2$, respectively (that is, the excess storage overhead is at least $1/r$ or $1$, respectively). 
\end{abstract}


%
\IEEEpeerreviewmaketitle

\section{Introduction}
%
%
%
A study sponsored by the storage company EMC found that the world's data is doubling every two year, and estimated it at 1.8 zettabytes (1.8 trillion gigabytes) in 2011.
\footnote{http://www.emc.com/about/news/press/2011/20110628-01.htm} Given these enormous volumes, the importance of efficient data storage can hardly be overestimated. 
%
These huge amounts of data need to be stored and reliably maintained over time, while being stored on individually unreliable components. 
To guarantee data survival over time, redundancy must be introduced. In distributed 
storage systems (DSS), typically data objects are stored in encoded form onto multiple storage units or {\em storage nodes\/}. In older 
DSS,
data blocks were simply replicated, but the actual, enormous  scale of operations demands the use of more sophisticated erasure coding techniques. 
%
Currently,  Reed-Solomon codes and other erasure codes are employed in cloud environments like  
Microsoft 
Windows Azure Storage \cite{WAS}, and in peer-to-peer storage systems like Wuala, Cleversafe, Oceanstore, and TotalRecall, see e.g., \cite{nc-survey}, \cite{overview-ddsc}, and references therein.


The use of erasure codes potentially affords orders of magnitude greater reliability while requiring less storage overhead, but to achieve this potential, it is of crucial importance  to find  efficient solutions for the {\em repair problem\/},
 the problem of maintaining system reliability in the presence of node failures.
Over time, storage nodes will leave the system due to node failures, caused for example by hardware failures (i.e., disk failures) or software updates in data centers, or peer churning in peer-to-peer systems. Under the simplest and most straightforward repair regime called {\em exact repair\/}, 
each data block stored on a failed node 
has to be exactly reconstructed and stored on a {\em newcomer node\/}.
In a more subtle repair regime called {\em functional repair\/}, we do not require that the newcomer stores an exact copy of the  lost data block, but 
typically the data block stored in the newcomer node will be some linear combinations of the data blocks in the other nodes, not necessarily exactly equal to the lost data block but enabling recovery of the originally stored information in combination with the data blocks on the other nodes (later, we will discuss an example).

Various performance metrics for repair efficiency have been considered. The total amount of information communicated during repair (called the repair bandwidth \cite{Dimakis}) has received the most attention,  and is currently best understood. However, for certain applications like cloud storage and deep archival minimizing disk I/O seems more valuable \cite{kbp:11:iso}. Since the disk I/O is proportional to the  number of nodes contacted during repair of a failed node, the repair locality of a storage code has recently emerged as an important parameter.

In this paper, we first present a brief overview of the cutset bound and regenerating codes from \cite{Dimakis}, discussing various types of storage codes along the way. Then we review the recent work on repair locality and present some new results. We end by suggesting some  directions for further research.  

For a general, more complete overview of DDS and storage codes, we refer to \cite{nc-survey}, \cite{overview-ddsc}, and to the Storage Wiki \cite{StorageWiki}.


\section{\label{Sreg}Regenerating codes}
Assume that a data object is stored in encoded form across $n$ storage nodes of a DSS, with each of the nodes storing one {\em data block\/}, an amount $\ga$ of data of the encoded object.
%
%
%
%
%
%
When a node fails, a newcomer node is allowed to contact {\em any\/} set of $r$ live nodes and to download an amount $\gb$ of data from each of them in order to {\em regenerate\/} some of the lost
data, in the form of a {\em replacement block\/}, again containing an amount $\ga$ of data. This number $r$ is referred to as the {\em repair locality\/} or the {\em fan-in\/} of the repair process. (Note that in many earlier publications the letter $d$ is used instead.) We require, and this is essential, that this regeneration process ensures that a data collector can reconstruct the original data object, at any time during this process, from any $k$ of the resulting data blocks in the current $n$ live nodes, for some number $k$.
%
In what follows, we assume that $k$ is the {\em smallest\/} number with this property; note that then $k\leq r\leq n-1$.
Now the question that arises is: how much information can be stored given these assumptions?

The repair problem can be abstracted in terms of an information flow network, where a new node~$v$, having storage capacity $\ga$, is represented by 
a capacitated edge 
$v^{\rm in}\stackrel{\ga}{\longrightarrow} v^{\rm out}$ 
and with $r$ capacitated edges $w_i^{\rm out} \stackrel{\gb}{\longrightarrow} v^{\rm in}$, 
each of  of capacity $\gb$, 
representing the data flow from the nodes assisting in the repair  towards the new node during regeneration.
Now the problem is reduced to a multicasting problem on this network. Network flow theory can be used to investigate the maximum possible flow of information towards a set of $k$ nodes used for data recovery by looking for possible bottlenecks in the network.
In the breakthrough paper \cite{Dimakis},  it is shown by such a a maxflow-mincut argument that the maximum amount $m$ of information that can be stored 
satisfies the cutset bound 
\beql{Ecutset} 
m\leq \sum_{j=0}^{k-1}  \min\{(r-j)\gb, \ga).
\eeql

Storage codes for this model that meet 
the above bound are called {\em Regenerating Codes\/}.  Two types of Regenerating Codes are of special interest, one corresponding to the point of optimal storage efficiency and the other to the point of optimal repair bandwidth efficiency. Since any $k$ nodes contain all available information, nodes must have storage capacity $\ga\geq m/k$. Regenerating  Codes with $\ga=m/k$  minimize the required amount of storage among regenerating $(n,k,r)$ codes; such codes are called Minimum Storage Regenerating (MSR) codes. They are characterized by having $\ga=m/k=(r-k+1)\gb$. On the other hand, a data amount of at most $\gc=r\gb$  is available during repair of a node, so that $r\gb\geq \ga$. Code with $r\gb=\ga$ minimize the repair bandwidth $\gc=r\gb$ among regenerating $(n,k,r)$ codes; such codes are called Minimum Bandwidth Regenerating (MBR) codes. They are characterized by having $\ga=r\gb$ and $m/\gb=kr-{k\choose 2}$.

The existence of Regenerating Codes (and even of linear ones) for all feasible parameter sets (i.e., with $n>r\geq k$, assuming the value for $k$ is minimal) essentially follows from results in Network Coding \cite{Dimakis}, but explicit constructions are not immediately available, nor are they obvious.  Moreover, to make matters worse, strictly speaking the said results only guarantee existence of these codes {\em for functional repair\/} of a number of node failures that is {\em bounded\/} over time, if a {\em sufficiently large\/} field is employed. 
It would be hard to imagine that the boundedness condition is really essential, and indeed it has been lifted, first for $r=n-1$ in \cite{Wu07deterministicregenerating}, and later for all parameter sets on the cutset bound in \cite{5402495}. The resulting codes do prove existence, but are not practical.

Fortunately, for the important cases of MSR and MBR codes, as well as for some other cases, explicit constructions are now known for functional repair, and in many cases also for exact repair. Before we 
review 
these results, we 
discuss a useful abstract description of storage codes for 
exact and functional repair,
and provide
some examples.

\section{Linear storage codes}
Recall that, under the exact repair regime, 
each data block on a failed storage node has to be exactly reconstructed and stored on a newcomer node.
Just as linear error-correcting or erasure-correcting codes are best thought of simply as vector spaces over a finite field, whose properties relative to notions like (Hamming) distance can then be studied, we believe that linear distributed storage codes for the exact-repair regime are best thought of in a similar way, now as a collection of subspaces of a fixed vector space, for which then similar appropriate notions can be introduced and investigated. So we will first introduce them in this way, along with various relevant notions. Then, we will explain  how to use a storage code to actually store and maintain information, and discuss some examples to illustrate the concepts. 
Our approach should be compared to the one as found for example  in \cite{DBLP:journals/tit/ShahRKR12a} or \cite{5714273}.

\subsection{Exact-repair storage codes as collections of vector spaces}
A {\em linear exact-repair distributed storage code (LERSC)\/} $\cU$, 
with parameters $(m; n, \ga)$ over a (finite) field~$\bbF$ is a collection of~$n$ subspaces $U_1, \ldots, U_n$ of an $m$-dimensional vector space~$U$ over~$\bbF$, each of dimension~$\ga$. We will refer to the space $U$ as the {\em message space\/} and to the $U_i$ as the {\em storage spaces\/}. The integer $\ga$ is called the {\em storage capacity\/}.

A subset $K$ of the storage spaces is called a {\em recovery set\/} of the storage code $\cU$ if these subspaces together span the entire vector space~$U$.
Here, the {\em span\/} of a collection of vector spaces $A_1, \ldots, A_h$  is the collection of all vectors $a_1+\cdots a_h$ with $a_i\in A_{i}$ for all $i$, that is, the smallest vector space containing all the vector spaces $A_1, \ldots, A_h$.
The {\em  recovery dimension\/} $k=k(\cU)$ of $\cU$ is defined as the size of the {\em smallest\/} recovery set of~$\cU$. 

Given some positive integer~$\gb$, referred to as the {\em transport capacity\/}, we say that 
a collection $R$ of subspaces of the storage code $\cU$ is a {\em repair set \/} for a certain subspace $U_\ell\notin R$  if it is possible to choose a $\gb$-dimensional {\em repair space\/} $W_{i,\ell}\subset U_i$ for each $U_i\in R$ such that $U_\ell$ is contained in the span of the repair spaces $W_{i,\ell}$.
If each subspace in the storage code has a repair set  of size~$r$ w.r.t.\ transport capacity $\gb$ then we say that the code $\cU$ has {\em repair locality\/} $r$ with respect to transport capacity $\gb$. We will refer to a storage code $\cU$ with all the above parameters as an  $(m; n,k,r, \ga,\gb)$-storage code. 

Now let us see how such a storage code $\cU$ can be used to actually store and maintain information. The information to be stored will be represented by a vector $x\in U$. So, for example, if  $\bbF$ has size $q=2^h$, then $x$ represents a file consisting of $mh$ bits, grouped into $m$ symbols of $h$ bits each. 
Now consider a DSS consisting of~$n$ storage units or {\em storage nodes\/} $v_1, \ldots, v_n$.  
In each subspace $U_i$ of $\cU$ we choose a basis $b_{i,1}, \ldots, b_{i,\ga}$, represented by the $\ga\times m$ matrix $B_i=[b_{i,1}\cdots b_{i,\ga}]$. 
Then, we associate the subspace $U_i$ with storage node $v_i$, and use this node to store the $\ga$ symbols of the vector $B^\top_ix$, that is, in $v_i$ we store the $\ga$ inner products of $x$ with the basis vectors of $U_i$. Using only simple linear algebra, it is easily seen that indeed a data collector can recover the vector $x$ by collecting the 
set of vectors $B_i^\top x$ stored in a subset $K$ of the nodes 
 if (and only if)  these nodes constitute a recovery set of~$\cU$. (Here it is of course assumed that the choice of the matrices $B_i$ is known to the data collector.)


Similarly, given a repair set $R$ for a node $v_\ell$ w.r.t.\ transport capacity $\gb$, we choose a fixed basis in each repair space $W_{i,\ell}$ inside subspace $U_i\in R$, represented by  a $\gb\times \ga$ {\em repair\/} matrix $T_{i,\ell}$ having this basis as columns. Again, it is easily seen that (a)  each node $v_i\in R$ can compute $T_{i,\ell}^\top x$ from the vector $B_i^\top x$ stored in $v_i$ and (b) node $v_\ell$ can recompute the vector $B_\ell^\top x$ from the vectors $T_{i,\ell}^\top x$ collected during repair from the nodes in the repair set $R$.
(Here, we assume that the choice of the repair matrices $T_{i,\ell}$ is known in node $v_\ell$.) 


Note that a storage code as above has  a coding rate $R(\cU)=m/(n\ga)$ and excess storage overhead $o(\cU) =1/R(\cU)-1=(n\ga-m)/m$. 

\bex{E42} Consider the storage code $\cU=\{U_0,U_1,U_2,U_3\}$ over the binary field $\bbF_2$ with $U_0=\langle e_0, e_2+e_3\rangle$, $U_1=\langle e_1, e_3+e_0\rangle$, $U_2=\langle e_2, e_0+e_1\rangle$, and $U_3=\langle e_3, e_1+e_2\rangle$, considered as subspaces of $U=\bbF_2^4$. Here, we write $\langle a_1, \ldots, a_h\rangle$ to denote the span of the vectors $a_1,\ldots, a_h$, the vector space consisting of all linear combinations of $a_1,\ldots, a_h$. We claim that the code $\cU$ is an $(m=4; n=4,k=2,r=3,\ga=2,\gb=1)$ linear exact-repair storage code (LERSC). Indeed, there are $n=4$ subspaces $U_i$, each of dimension $\ga=2$. 
Furthermore, $U$ has dimension $m=4$, and $k=2$ since 
any two subspaces intersect trivially, so together span~$U$.
To repair node $v_0$ using the size $r=3$ repair set $R=\{U_1, U_2, U_3\}$,  we choose repair spaces $W_{1,0}=\langle e_0+e_3\rangle \subseteq U_1$, $W_{2,0}=\langle e_2\rangle \subseteq U_2$, and $W_{3,0}=\langle e_3\rangle \subseteq U_3$, each of dimension $\gb=1$. Note that this choice is valid since indeed $U_0\subseteq \langle e_0+e_3,e_2, e_3\rangle$. The storage code $\cU$ is invariant under the linear transformation given by $e_i\mapsto e_{i+1}$ (indices modulo 3), so the repair spaces for other nodes can be obtained by symmetry. With the bases as suggested by the above description, this code stores a vector $x=(x_0, \ldots, x_3)$ by letting node 0 hold $x_0$ and $x_2+x_3$,
and repairs node 0 by downloading $x_0+x_3$ from node 1, $x_2$ from node 2, and $x_3$ from node 3. 
\IEEEQED
\eex
 
A linear transformation fixing  the storage code such as the cyclic shift in the example above could be termed an {\em automorphism\/} of the code.
The notion of code automorphisms has been very fruitful in the field of error-correcting codes, where it has lead to the discovery of several important classes of codes such as cyclic codes, of which Reed-Solomon codes are a special case. But in contrast, until now symmetry has not played a significant role in storage codes. It might be of interest to systematically search for storage code with extra symmetries.

A LERSC $\cU$ is essentially determined by the subspaces contained in~$\cU$, however, as seen above the {\em actual implementation\/} of the code also depends on the choice of bases in the various spaces. This choice can have a crucial influence on the performance of the code. Ideally, each repair subspace is spanned by a subset of the basis in the node; in that case, during repair each node simply transfers a {\em subset\/} of its data, so that no computations are required. This situation, referred to as {\em repair-by-transfer\/}, is illustrated below.

%
\bex{Erate1/2} 
We construct a simple binary rate-(1/2) repair-by-transfer $(m={n\choose 2}; n, k=n-1,r=n-1, \ga=n-1, \gb=1)$ storage code.
(In fact, these codes are MBR codes.)
The message space $U$ has dimension $m={n \choose 2}$, 
so we can index the coordinate positions with pairs $\{i,j\}\subseteq \{1, \ldots, n\}$. Given a message vector $x=\{x_{\{i,j\}}\}$, we let node $v$ store the $\ga=n-1$ symbols $x_{\{v,j\}}$ ($j\neq v$). If node $v$ fails, it can be exactly repaired by downloading symbol $x_{\{v,j\}}$ from node $j$, for each node $j\neq v$. In other words, the node subspaces are $U_v=\langle e_{\{v,j\}}\mid j\neq v\rangle$, with repair spaces $W_{j,\ell}=\langle e_{\{j,\ell\}}\rangle$.
\IEEEQED
\eex

The Fractional Repetition Codes described in \cite{allerton-fractional} combine a repair-by-transfer inner code with an MDS outer code (that is, the stored vector $x$ in the message space is itself a codeword in an MDS code); these codes actually meet the cutset bound (\ref{Ecutset}) at the MBR point.

\subsection{Linear functional-repair storage codes}
Under the regime of functional repair, a data block on a failed storage node has to be replaced by a data block on a newcomer that is {\em information equivalent\/} to the one on the failed node, while ensuring the possibility of future functional repair of other nodes. Linear distributed storage codes for functional repair are perhaps best thought of as a {\em specification\/} of a subspace arrangement, with the property that in any realization, a subspace can be ``repaired'' by replacing it with a (possibly different) subspace so that the resulting arrangement again satisfies the specifications. An example will help to illustrate the idea.
\bex{MBR-func}
We will construct a linear functional-repair storage code $\cU$ with parameters $(m=5; n=4, k=r=3, \ga=2, \gb=1)$,  so with coding rate $R=5/8$.
Note that this parameter set meets the cutset bound (\ref{Ecutset}),  in a point different from the MBR and MSR points.

Let $U$  be a 5-dimensional vector space over $\bbF_2$. We will ensure that at each moment in time, the four 2-dimensional  storage subspaces $U_1, \ldots, U_4$ associated with the four storage nodes comply with the following specification:
\begin{enumerate}
\item Any two of the storage spaces intersect trivially, that is, $U_i\cap U_j=\{0\}$ when $i\neq j$;
\item Any three of the storage spaces span $U$. 
\end{enumerate}
Suppose that $U_1, \ldots, U_4$ satisfy these constraints, and suppose that node 4 fails. 
Without loss of generality, we may assume that  
$U_1=\langle e_1, e_3\rangle$, $U_2=\langle e_2, e_4\rangle$, and $U_3 = \langle e_5, e_1+e_2\rangle$,  for some basis $e_1, \ldots, e_5$ of~$U$. Indeed, $U_3$ must have trivial intersection with both $U_1$ and $U_2$, but, having dimension 2, necessarily intersects the 4-dimensional span $U_1+ U_2$, hence this intersection is of the form $e_1+e_2$ with $e_i\in U_i^*=U_i\setminus \{0\}$.  This shows that $U_1, U_2, U_3$ have the indicated form. Now, to repair (or initially construct) the storage space $U_4$, given that $\gb=1$ we must choose a vector $a_i\in U_i^*$ for $i=1,2,3$, and let $U_4$ be some 2-dimensional subspace of their span $\langle a_1, a_2, a_3\rangle$, which by rule 1 should  not contain any of the $a_i$.  Hence $U_4$ is of the form $\{0, a_1+a_2, a_1+a_3, a_2+a_3\}$. Finally, $a_3\neq e_1+e_2$ since otherwise $U_4\subset U_1+U_2$, violating rule 2, and similarly, $a_1\neq e_1, a_2\neq e_2$. So $a_1=e_3+x_1 e_1, a_2=e_4+x_2e_2, a_3=e_5+x_3(e_1+e_2)$, and it is now easily verified that any choice of $x_1, x_2, x_3\in \bbF_2$ is valid. 
(Initially, we can take for example $U_4=\langle e_3+e_4, e_3+e_5\rangle$.) This shows that we can maintain the specification forever, provided that never two nodes fail at the same time.
\IEEEQED
\eex
The use of functional-repair storage codes as above to actually store information is similar to that of the exact-repair storage codes introduced earlier, except that now at each moment the other nodes and the data collector have to be informed of the actual {\em state\/} of a storage node, that is, of its current  storage space. This extra overhead can be relatively small if the code is used 
to store a large number of messages {\em simultaneously\/}.

\subsection{Existence of regenerating storage codes on the cutset bound}
We end this section with a brief overview of the known constructions and nonexistence results to date. As mentioned before, regenerating codes have been shown to exist for all parameter sets on the cutset bound (\ref{Ecutset}).

For the MBR point (minimizing repair bandwidth), linear exact-repair regenerating storage codes have been  constructed for all parameter sets in \cite{DBLP:journals/tit/RashmiSK11} using a product-matrix construction, with a field size of the order of the number $n$ of nodes.  Exact-repair-by-transfer regenerating  MBR codes have been constructed for the case $r=n-1$,  now using field sizes of order $n^2$ \cite{6062413}. 
 
Exact-repair MSR regenerating storage codes have been constructed for all parameter sets with $r\geq 2k-2$ in \cite{DBLP:journals/tit/RashmiSK11} 
(for some other constructions in this range, see the references on the Storage Wiki \cite{StorageWiki});
the non-existence of exact-repair regenerating MSR codes with $r<2k-2$ for the case $\gb=1$ (commonly referred to as  ``no symbol extension'') was demonstrated in {\cite{DBLP:journals/tit/ShahRKR12a}, by showing that a phenomenon called {\em interference alignment\/} necessary must occur in such codes. To complete the picture,   \cite{DBLP:journals/corr/abs-1004-4663} and \cite{5507931} have shown asymptotic existence of exact-repair regenerating MSR storage codes for all $n,k,r$ (that is, for points arbitrarily close to the cutset bound, for sufficiently large file sizes).
Finally, functional repair-by-transfer regenerating MSR codes for parameter sets with $k=2$ and $r=n-1$ have been constructed in \cite{shum-hu-func}.

The paper \cite{6062413} also shows the non-achievability of essentially all {\em interior points\/} on the cutset bound (that is, different from MBR and MSR) for exact repair in the case $\gb=1$ (no symbol extension).















\section{Repair locality in storage codes}
Application contexts like cloud storage systems and deep archival storage require  a low disk I/O overhead \cite{kbp:11:iso}. 
Since the disk I/O is proportional to the number of nodes involved in a repair, this makes
the repair locality an important performance metric, 
which was recognized 
in
\cite{DBLP:conf/infocom/OggierD11}, 
\cite{DBLP:journals/eccc/GopalanHSY11}, \cite{DBLP:conf/infocom/PapailiopoulosLDHL12}.  Codes designed for small repair locality are for example Pyramid codes \cite{4276609}, Homomorphic codes \cite{DBLP:conf/infocom/OggierD11} and Spread codes \cite{6089443},  codes in \cite{DBLP:conf/infocom/PapailiopoulosLDHL12}, and LRC codes \cite{DBLP:journals/corr/abs-1204-6098}. Some of the repair-by-transfer codes in \cite{allerton-fractional} and \cite{6033980} can also be considered as designed for this purpose. 

Already in \cite{ kbp:11:iso}, it was conjectured that there are trade-offs  between recovery I/O and storage efficiency. Up to now, bounds have been developed in the case of exact repair,  involving rate, repair locality, and code distance.
For linear $[n,k,d]$-codes, it was shown in  \cite{6259860} that $n-k \geq \lceil k/r\rceil +d-2$ (attainable for $d\geq 2$), implying that the rate $R=k/n$ satisfies $R\leq r/(r+1)$. A more general  information-theoretical bound derived in~\cite{lrc} (see also its full version \cite{lrc-scworl}), states that $d\leq n-\lceil  m/\ga\rceil -  \lceil m/(r\ga)\rceil +2$, where $m$ is the amount of encoded information, $d$ the ``information-theoretical distance'' of the code (defined as the maximum number such that any $k=n-d+1$ nodes can reconstruct the stored information) and $\ga$ the storage per node;  in the case where $(r+1)|n$, a code was constructed with $d=  n-\lceil m/\ga\rceil -  \lceil m/(r\ga)\rceil -1$. 
Again,   if any failed node can be repaired at all then $d\geq2$, in which case  the bound implies that the rate $R=m/(n\ga)\leq r/(r+1)$.

In all the models discussed above, a given node and all its reincarnations are assumed to have the same, {\em fixed\/} repair set of size~$r$.  
Our aim is to investigate the trade-off between rate and repair locality in 
an information flow network  setting similar to that of the cutset bound (\ref{Ecutset}). 
Remark that the cutset bound does not depend on the requirement that {\em every\/} set of $k$ nodes can recover the stored information: indeed, inspection of the proof in \cite{Dimakis} shows that the cutset bound still holds if we only assume that {\em some\/} set of $k$ nodes has this property, as long as a newcomer node can connect to {\em any\/} set of $r$ live nodes during repair. Already in \cite{Dimakis}, the question was raised if the mincut value could be larger if a newcomer could {\em choose\/} the $r$ live nodes to connect to. It is precisely this question that we investigate here.


So assume that we have a storage code for the functional repair regime that can store a total amount $m$ of information by storing an amount $\ga$  of data onto $n$ nodes, with the further property that at all times, a failed node can be (functionally) repaired by downloading  from each member of some set of $r$ nodes an amount $\gb$ of data, so that at any time during this ongoing process the original information can be fully retrieved. Then what can be said about the maximum coding rate $R=m/(n\ga)$? 
The question can be formulated in terms of a game played by two players, KILLER and BUILDER, on the information flow graph as in \cite{Dimakis}. Originally, the graph consists of $n$ isolated live nodes. The two players move in turn; KILLER moves by choosing a node and killing it, then BUILDER moves by creating a new live node and connecting to it from some set of $r$ live nodes of his choice. 
The aim of KILLER is to force a cutset of {\em small\/} capacity, and BUILDER tries to prevent that. Remark that the maximum amount of information that can be maintained in the storage system is at most equal to the capacity of any cutset at any stage of the game. The result of the game under optimal play by both players thus provides an upper bound on $m$.
In \cite{hdlh-maxrategen}, we use this game to prove the  following results.
\btm{Tmax} With the above notation and assumptions, we have the following.
\begin{enumerate}
\item If $\ga=\gb$, then $R\leq r/(r+1)$. Equality holds for (exact-repair) MDS codes with $n=r+1$.
\item If $\ga=r\gb$, then $R\leq 1/2$. Equality holds for the exact-repair-by-transfer linear storage codes in Example~\ref{Erate1/2}.
\end{enumerate}
\etm
\btm{Tmax2n} With the same notation and assumptions, for $r=2$ we have that 
%
\[R \leq \frac{\ga+\gb}{3\ga}.\] 
More precisely, if $n=3q-e$ with $e\in \{0,1,2\}$, then
\[m \leq q \ga +(q-e)\gb.\] 
\etm
Note that the examples mentioned in Theorem~\ref{Tmax} allow the construction of codes of length $n=r+1$ attaining the bounds in all cases mentioned in the above theorems, as well as  construction of optimal (repetition) codes of lengths $n$ whenever $r+1|n$.















Recently \cite{shum-hu-coop}, \cite{5962548},  \cite{5978920}, generalizations of the cutset bound from \cite{Dimakis} have been derived  in an information flow network setting similar to the one  in Section~\ref{Sreg}, now for the case where a number $s$ of nodes is repaired {\em simultaneously\/}. Here, during repair each of the $s$ newcomer nodes is allowed to download an amount $\gb_1$ of data from a set of live nodes of size~$r$,  and subsequently an amount $\gb_2$ of data from each of the {\em other\/} newcomer nodes. It would be interesting to generalize our bounds to this more general setting.


\section{Conclusion}
We  have investigated the trade-off between the coding rate $R$  and repair locality $r$ in the functional repair regime, in a information flow network setting. Tight bounds are presented for the two extreme cases $\ga=\gb$, where $R\leq r/(r+1)$,  and $\ga=r\gb$, where  $R\leq 1/2$, and for the case where $r=2$. 

\section*{Acknowledgment}

The author would like to thank Lluis Pamies-Juarez and Fr\'ed\'erique Oggier for proofreading and for providing some helpful comments.
The research of Henk D.L.~Hollmann is supported by the Singapore National
Research Foundation under Research Grant
NRF-CRP2-2007-03.





\begin{thebibliography}{10}
\providecommand{\url}[1]{#1}
\csname url@samestyle\endcsname
\providecommand{\newblock}{\relax}
\providecommand{\bibinfo}[2]{#2}
\providecommand{\BIBentrySTDinterwordspacing}{\spaceskip=0pt\relax}
\providecommand{\BIBentryALTinterwordstretchfactor}{4}
\providecommand{\BIBentryALTinterwordspacing}{\spaceskip=\fontdimen2\font plus
\BIBentryALTinterwordstretchfactor\fontdimen3\font minus
  \fontdimen4\font\relax}
\providecommand{\BIBforeignlanguage}[2]{{%
\expandafter\ifx\csname l@#1\endcsname\relax
\typeout{** WARNING: IEEEtran.bst: No hyphenation pattern has been}%
\typeout{** loaded for the language `#1'. Using the pattern for}%
\typeout{** the default language instead.}%
\else
\language=\csname l@#1\endcsname
\fi
#2}}
\providecommand{\BIBdecl}{\relax}
\BIBdecl

\bibitem{WAS}
H.~C.~Huang, Y.~X. Simitci, A.~Ogus, B.~Calder, P.~Gopalan, J.~Li, and
  S.~Yekhanin, ``Erasure coding in windows azure storage.''\hskip 1em plus
  0.5em minus 0.4em\relax USENIX ATC, Boston, MA, June 2012.

\bibitem{nc-survey}
A.~G. Dimakis, K.~Ramchandran, Y.~Wu, and C.~Suh, ``A survey on network codes
  for distributed storage,'' \emph{IEEE Proceedings}, vol.~99, no.~3, pp.
  476--489, March 2011.

\bibitem{overview-ddsc}
\BIBentryALTinterwordspacing
A.~Datta and F.~Oggier. (2011) An overview of codes tailor-made for networked
  distributed data storage. [Online]. Available:
  \url{http://arxiv.org/abs/1109.2317}
\BIBentrySTDinterwordspacing

\bibitem{Dimakis}
A.~G. Dimakis, P.~B. Godfrey, Y.~Wu, M.~Wainwright, and K.~Ramchandran,
  ``Network coding for distributed storage systems,'' \emph{{IEEE} Trans. Inf.
  Theory}, vol.~56, no.~9, 2010.

\bibitem{kbp:11:iso}
O.~Khan, R.~Burns, J.~S. Plank, and C.~Huang, ``In search of {I/O}-optimal
  recovery from disk failures,'' in \emph{HotStorage '11: 3rd Workshop on Hot
  Topics in Storage and File Systems}.\hskip 1em plus 0.5em minus 0.4em\relax
  Portland: USENIX, June 2011.

\bibitem{StorageWiki}
\BIBentryALTinterwordspacing
``The erasure coding for distributed storage wiki.'' [Online]. Available:
  \url{http://tinyurl.com/storagecoding}
\BIBentrySTDinterwordspacing

\bibitem{Wu07deterministicregenerating}
Y.~Wu, R.~Dimakis, and K.~Ramchandran, ``Deterministic regenerating codes for
  distributed storage,'' in \emph{In Allerton Con.\ Control, Computing, and
  communication, Urbana-Campaign, IL}, September 2007.

\bibitem{5402495}
Y.~Wu, ``Existence and construction of capacity-achieving network codes for
  distributed storage,'' \emph{Selected Areas in Communications, IEEE Journal
  on}, vol.~28, no.~2, pp. 277 --288, february 2010.

\bibitem{DBLP:journals/tit/ShahRKR12a}
N.~B. Shah, K.~V. Rashmi, P.~V. Kumar, and K.~Ramchandran, ``Interference
  alignment in regenerating codes for distributed storage: Necessity and code
  constructions,'' \emph{{IEEE} Trans. Inf. Theory}, vol.~58, no.~4, pp.
  2134--2158, 2012.

\bibitem{5714273}
C.~Suh and K.~Ramchandran, ``Exact-repair mds code construction using
  interference alignment,'' \emph{{IEEE} Trans. Inf. Theory}, vol.~57, no.~3,
  pp. 1425 --1442, march 2011.

\bibitem{allerton-fractional}
S.~Y.~E. Rouayheb and K.~Ramchandran, ``Fractional repetition codes for repair
  in distributed storage systems,'' in \emph{In Allerton Con.\ Control,
  Computing, and communication, Urbana-Campaign, IL}, 2010.

\bibitem{DBLP:journals/tit/RashmiSK11}
K.~V. Rashmi, N.~B. Shah, and P.~V. Kumar, ``Optimal exact-regenerating codes
  for distributed storage at the msr and mbr points via a product-matrix
  construction,'' \emph{{IEEE} Trans. Inf. Theory}, vol.~57, no.~8, pp.
  5227--5239, 2011.

\bibitem{6062413}
N.~Shah, K.~Rashmi, P.~Vijay~Kumar, and K.~Ramchandran, ``Distributed storage
  codes with repair-by-transfer and nonachievability of interior points on the
  storage-bandwidth tradeoff,'' \emph{{IEEE} Trans. Inf. Theory}, vol.~58,
  no.~3, pp. 1837 --1852, march 2012.

\bibitem{DBLP:journals/corr/abs-1004-4663}
C.~Suh and K.~Ramchandran, ``On the existence of optimal exact-repair mds codes
  for distributed storage,'' \emph{CoRR}, vol. abs/1004.4663, 2010.

\bibitem{5507931}
V.~R. Cadambe, S.~A. Jafar, and H.~Maleki, ``Minimum repair bandwidth for exact
  regeneration in distributed storage,'' in \emph{Wireless Network Coding
  Conference (WiNC), 2010 IEEE}, june 2010, pp. 1 --6.

\bibitem{shum-hu-func}
\BIBentryALTinterwordspacing
K.~Shum and Y.~Hu. Functional-repair-by-transfer regenerating codes. Will be
  presented at ISIT 2012. [Online]. Available:
  \url{home.ie.cuhk.edu.hk/~wkshum/papers/FRBT.pdf}
\BIBentrySTDinterwordspacing

\bibitem{DBLP:conf/infocom/OggierD11}
F.~E. Oggier and A.~Datta, ``Self-repairing homomorphic codes for distributed
  storage systems,'' in \emph{INFOCOM}.\hskip 1em plus 0.5em minus 0.4em\relax
  IEEE, 2011, pp. 1215--1223, extended version at
  http://arxiv.org/abs/1107.3129.

\bibitem{DBLP:journals/eccc/GopalanHSY11}
P.~Gopalan, C.~Huang, H.~Simitci, and S.~Yekhanin, ``On the locality of
  codeword symbols,'' \emph{Electronic Colloquium on Computational Complexity
  (ECCC)}, vol. 100, 2011, accepted for publication in {IEEE} Trans.\ Inform.\
  Theory.

\bibitem{DBLP:conf/infocom/PapailiopoulosLDHL12}
D.~S. Papailiopoulos, J.~Luo, A.~G. Dimakis, C.~Huang, and J.~Li, ``Simple
  regenerating codes: Network coding for cloud storage,'' in \emph{INFOCOM},
  A.~G. Greenberg and K.~Sohraby, Eds.\hskip 1em plus 0.5em minus 0.4em\relax
  IEEE, 2012, pp. 2801--2805.

\bibitem{4276609}
C.~Huang, M.~Chen, and J.~Li, ``Pyramid codes: Flexible schemes to trade space
  for access efficiency in reliable data storage systems,'' in \emph{Network
  Computing and Applications, 2007. NCA 2007. Sixth IEEE International
  Symposium on}, july 2007, pp. 79 --86.

\bibitem{6089443}
F.~Oggier and A.~Datta, ``Self-repairing codes for distributed storage -- a
  projective geometric construction,'' in \emph{Information Theory Workshop
  (ITW), 2011 IEEE}, oct. 2011, pp. 30 --34.

\bibitem{DBLP:journals/corr/abs-1204-6098}
A.~S. Rawat and S.~Vishwanath, ``On locality in distributed storage systems,''
  \emph{CoRR}, vol. abs/1204.6098, 2012.

\bibitem{6033980}
S.~Pawar, N.~Noorshams, S.~El~Rouayheb, and K.~Ramchandran, ``Dress codes for
  the storage cloud: Simple randomized constructions,'' in \emph{Information
  Theory Proceedings (ISIT), 2011 IEEE International Symposium on}, 31
  2011-aug. 5 2011, pp. 2338 --2342.

\bibitem{6259860}
P.~Gopalan, C.~Huang, H.~Simitci, and S.~Yekhanin, ``On the locality of
  codeword symbols,'' \emph{{IEEE} Trans. Inf. Theory}, vol.~PP, no.~99, p.~1,
  2012.

\bibitem{lrc}
\BIBentryALTinterwordspacing
D.~S. Papailiopoulos and A.~G. Dimakis. Locally repairable codes. Will be
  presented at ISIT 2012. [Online]. Available:
  \url{http://arxiv.org/abs/1206.3804}
\BIBentrySTDinterwordspacing

\bibitem{lrc-scworl}
\BIBentryALTinterwordspacing
------. Storage codes with optimal repair locality. [Online]. Available:
  \url{http:\\tinyurl.com/82cucvd}
\BIBentrySTDinterwordspacing

\bibitem{hdlh-maxrategen}
H.~D.~L. Hollmann, ``On the minimum storage overhead of distributed storage
  codes with given repair locality,'' in preparation.

\bibitem{shum-hu-coop}
\BIBentryALTinterwordspacing
Y.~H. Kenneth W.~Shum. Cooperative regenerating codes. [Online]. Available:
  \url{http://arxiv.org/abs/1207.6762}
\BIBentrySTDinterwordspacing

\bibitem{5962548}
K.~Shum, ``Cooperative regenerating codes for distributed storage systems,'' in
  \emph{Communications (ICC), 2011 IEEE International Conference on}, june
  2011, pp. 1 --5.

\bibitem{5978920}
A.-M. Kermarrec, N.~Le~Scouarnec, and G.~Straub, ``Repairing multiple failures
  with coordinated and adaptive regenerating codes,'' in \emph{Network Coding
  (NetCod), 2011 Int.\ Symp.\ on}, July 2011, pp. 1 --6.

\end{thebibliography}




\end{document}